\documentclass[twocolumn,showpacs,preprintnumbers,amsmath,amssymb]{revtex4}

\usepackage{graphicx}
\usepackage{dcolumn}
\usepackage{bm}

\newcommand{\TEM}{${\rm TEM_{01}}\,$}

\begin{document}

\preprint{APS/123-QED}

\title{Bose Einstein Condensate in a Box}

\author{T.P. Meyrath}
\author{F. Schreck}
 \altaffiliation[Present address at ]{Institut f\"{u}r Quantenoptik und Quanteninformation, Innsbruck, Austria.}
\author{J.L. Hanssen}
 \altaffiliation[Present address at ]{National Institute of Standards and Technology, Gaithersburg, MD 20899, USA.}
\author{C.-S. Chuu}
\author{M.G. Raizen}
\affiliation{%
Center for Nonlinear Dynamics and Department of Physics \\
The University of Texas at Austin, Austin, Texas 78712-1081, USA
}%

\date{\today}

\begin{abstract}
Bose-Einstein condensates have been produced in an optical box trap.
This novel optical trap type has strong confinement in two
directions comparable to that which is possible in an optical
lattice, yet produces individual condensates rather than the
thousands typical of a lattice. The box trap is integrated with
single atom detection capability, paving the way for studies of
quantum atom statistics.
\end{abstract}

\pacs{03.75.-b, 32.80.Pj, 39.25.+k}
\maketitle

The field of atom optics has now reached the stage where atom
statistics are becoming a central theme following parallel
developments in quantum optics where photon statistics play a
crucial role.  The first direct measurement of a second order atomic
correlation was reported for a beam of metastable neon
\cite{Yasuda}.  In that case, a chaotic state was measured due to
the high temperature of the beam.  However the situation becomes
much more interesting at low temperatures where quantum statistics
play a role.  For example, atomic Fock states may be produced by the
Mott insulator transition \cite{Jaksch} and with a Quantum Tweezer
for atoms \cite{Diener}.   Atomic spatial antibunching should occur
for a Tonks gas of bosons \cite{Kheruntsyan}. In a recent experiment
\cite{Paredes}, the latter was inferred from suppression of
three-body loss, but was not a direct measurement of atom
statistics.  It is clear that these are only a few examples of an
emerging theme in atom optics which is becoming increasingly
important. The experiments performed to date have all relied on the
use of optical lattices \cite{Greiner, Paredes, Kinoshita, Tolra}.
While this tool has yielded impressive results, it does not allow
single-site control and addressability. Instead of a single lower
dimensional condensate, many thousands are created in parallel. For
example, in the case of the Mott insulator transition, an ideal
measurement would be to turn off all sites except one and directly
record the single site atom statistics. Unfortunately this has not
been possible in an optical lattice. Likewise, for the predicted
Quantum Tweezer, a single quantum dot must be produced in order to
determine the exact atomic state.
\begin{figure}
\includegraphics{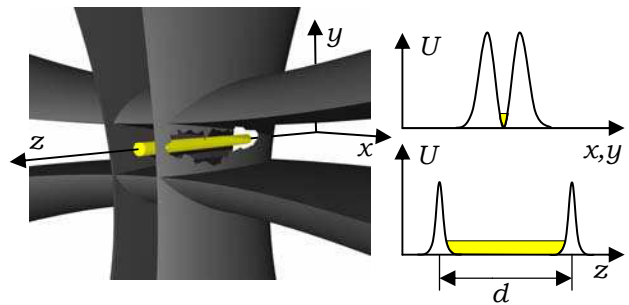}
\caption{\label{fig:BeamSetup} The crossed \TEM beams (x\TEM) are
shown pictorially on the left. The potential shapes of the trap are
given in arbitrary units on the right. The potential, $U$, in the
$x$ and $y$ directions have the shape of their respective \TEM
beams. The end-cap beams produce a trap along the $z$ axis, the
Gaussian walls are separated by $d=80\,\rm{\mu m}$. The end-caps are
not shown on the left, gravity is down in the pictorial.}
\end{figure}

Motivated by these goals, we have developed a new experimental
approach presented in this Rapid Communication. This system includes a novel
optical trap together with single-atom detection capability. Our
trap consists of a crossed pair of elongated Hermite-Gaussian \TEM
mode beams \cite{Saleh}: horizontal (h\TEM) and vertical (v\TEM)
supplemented by Gaussian beam end-caps.  The geometry is illustrated
pictorially in Fig. \ref{fig:BeamSetup}. With this setup we obtain
trap frequencies in two-dimensions which are comparable to those
typically reported for optical lattices, however there is only a
single condensate in one-dimension (1D).  The axial motion is
confined by optical end-caps, producing the textbook geometry of a
``particle in a box.''   The resulting atomic number in this box is
generally under 3500 and is controlled by evaporation timing and
spacing of the end-caps \cite{oned_bec}. Single atom detection with
nearly unit quantum efficiency has been demonstrated and is fully
integrated with the new trap, paving the way for direct measurements
of quantum atom statistics.

The key feature of our experiment is the optical trap. Work using
Laguerre-Gaussian (LG) optical traps has produced individual 1D
condensates \cite{Bongs}, but not with radial confinement sufficient
for experiments of the Mott-insulator \cite{Greiner} or quantum
tweezer \cite{Diener} sort. This geometry also is limited in trap
uniformity because small waist sizes result in a short Rayleigh
range which is along the axial trapping direction. Our trap geometry
intrinsically overcomes this limitation due to beam orientation.

\begin{figure}
\includegraphics{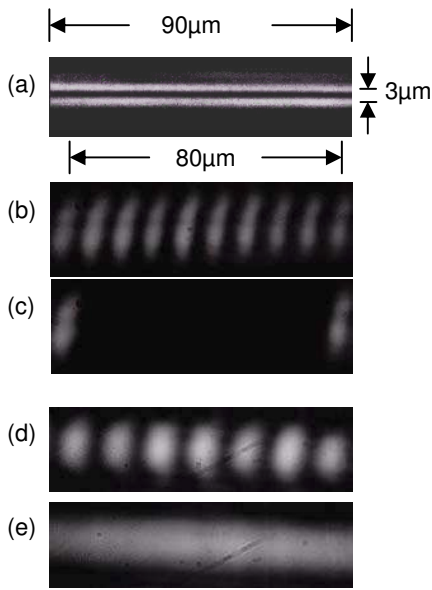}
\caption{\label{fig:beam_figure} Designer blue beams. Above are CCD
pictures of various beams imaged as on the atom cloud. (a) v\TEM
beam. (b) End-cap beam in a general setting showing 10 spots. (c)
End-cap beam as to end cap the x\TEM trap. (d) Compensation beam
driven with 7 frequencies. (e) Compensation beam with 80
frequencies. }
\end{figure}

Our rubidium 87 BEC experimental setup and method of producing \TEM beams is detailed in Ref. \cite{Meyrath}.
Fig. \ref{fig:beam_figure}(a)
shows a CCD image of the v\TEM beam. Both the v\TEM and h\TEM have a
waist radius of 125$\,\rm{\mu m}$ in the axial direction which is
much longer than the BEC region to provide a relatively uniform
trap. Theoretical values for trap depths and oscillation frequencies
are given in Table \ref{tab:beamtable} along with comparable
measured frequencies in the footnote. The measured values were
obtained by fitting time-of-flight expansions to those expected for
the harmonic oscillator ground states. These expansions show a
change of aspect ratio indicating that the phase transition to BEC
had occurred \cite{Gorlitz} with no thermal cloud observed.

\begin{table}
\caption{\label{tab:beamtable} Beams and parameters for the optical
trap. }
\begin{ruledtabular}
\begin{tabular}{lcccccc}
Beam\footnote{The blue beams are at $532\,\rm{nm}$, and the red
beams are at $1064\,\rm{nm}$, which produce repulsive and attractive
potentials, respectively \cite{Grimm}.} &
$w_{x,y}$\footnote{Measured radial beam waist, $x$ for vertical and
$y$ for horizontal beams.} & $w_z$\footnote{Measured beam waist in
axial direction.} &
   $P$\footnote{Measured beam power.} & $\omega/2\pi$\footnote{Calculated trap frequency in the radial
   direction. Measured values for full h\TEM and v\TEM are $24\pm4\,$kHz and
   $66\pm7\,$kHz.} & $
   U_0/k_B$\footnote{Calculated peak potential height/depth divided
   by the Boltzmann constant \cite{Grimm}.} \\
 & $(\rm{\mu m})$ & $(\rm{\mu m})$ & $(\rm{W})$ & $(\rm{kHz})$ & $(\rm{\mu K})$ & \\
\hline
h\TEM (Blue) & 2.4 & 125 &  &  &   \\
$\;\;$gavito-optical &  &  & 0.165 & 0.85 & 15  \\
$\;\;$weak trap &  &  & 0.165 & 8.3 & 15  \\
$\;\;$trap, evap. &  &  & 0.1 & 6.5 & 9.3  \\
$\;\;$full trap &  &  & 1.0 & 21 & 93   \\
\hline
v\TEM (Blue) & 1.8 & 125 &  &  &   \\
$\;\;$weak trap &  & & 0.74 & 27 & 92  \\
$\;\;$full trap &  &  & 3.7 & 61 & 460  \\
\hline
End-cap (Blue) & 6.1 & 2.5 & 0.011 &  & 28  \\
Compensation (Blue) & 9.8 & 7.2 & 0.001 &  & 0.54  \\
Vert. Circular\footnote{`Circular' and `Elliptical' refer to the
shape of the intensity profile of the beam, not the polarization.
All beams here are linearly polarized.\label{footnote}}
 (Red) & 50 & 50 & 0.021 & 0.056 & 0.8  \\
Vert. Elliptical$\,^g$ (Red)& 10 & 125 & 0.085 & 0.8 & 6.6  \\
\end{tabular}
\end{ruledtabular}
\end{table}
The preparation of atoms in a 1D box trap consists of several steps
as outlined here: (1) transfer the atoms into a combined
optical-magnetic trap, (2) produce a pancake shaped BEC by
evaporation while ramping off the magnetic trap, (3) transfer the
BEC into the h\TEM trap, (4) squeeze the BEC in another direction
with the elliptical red vertical trap, (5) load the elongated cloud
into the v\TEM trap and add end-caps, (6) remove the red beams and
add the compensation beam, (7) ramp up the x\TEM trap to full power.

\begin{figure}
\includegraphics{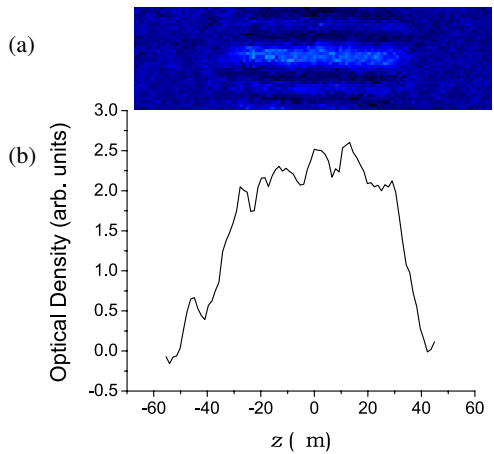}
\caption{\label{fig:becfigure} (a) An absorption image of a BEC of
$3\times 10^3$ atoms in a box with gaussian walls spaced by
$80\,\rm{\mu m}$ along the $z$ axis. (b) The profile of the BEC
along the $z$ axis integrated vertically. The image is {\it in
situ}, where the absorption beam is turned on in addition to the
optical trap. Resolution is limited by expansion during the
$30\,\rm{\mu s}$ exposure and the upper and lower stripes are
imaging artifacts. }
\end{figure}

The initial configuration in step (1) consists of the addition of a
horizontal blue sheet of light below the atoms in the 20$\,$Hz
magnetic trap along with the vertical circular red trap (`h\TEM,
gravito-optical' and `Vert. Circular' in Table \ref{tab:beamtable}). The
blue sheet is actually a \TEM mode beam which is located below the
magnetically trapped atoms. The atoms are initially above the beam
rather than in the node because the cloud is too large to be
captured directly from the magnetic trap. The center of the magnetic
trap is shifted downwards such that the atoms are pressed against
the sheet and the elastic collision rate is high enough for
evaporation. The magnetic trap is ramped off while the circular red
beam intensity is slightly lowered to allow for radial evaporation
resulting in a BEC of up to $3\times 10^5$ atoms. Gravity presses
the cloud into a pancake shape. In order to transfer the atoms into
the h\TEM beam, an upper h\TEM beam is ramped on in addition to and
$4\,\rm{\mu m}$ above the lower sheet h\TEM in $200\,\rm{ms}$. This
additional beam is a multiplex of the same beam (`h\TEM, weak trap'
in Table \ref{tab:beamtable}). It surrounds the pancake shaped BEC in
the vertical direction and the lower sheet is then removed. The
pancake shaped cloud must be compressed in another direction in
order to fit into the v\TEM beam, which is accomplished by ramping
up the elliptical and ramping down the circular vertical red traps.
This transfers the atoms into an elongated geometry and occurs in
two $100\,\rm{ms}$ stages. With the cloud elongated, the v\TEM beam
is ramped up in $100\,\rm{ms}$ to the weak value given in the table
and the end-cap beams are added spaced $80\,\rm{\mu m}$ apart along
the $z$ axis. The red trap is ramped off and the compensation beam
ramped on. This beam is used to improve the smoothness of the axial potential giving
it a box-like shape. The atoms are now in the weak x\TEM box, that
is, the atoms sit in a potential such that on axis there are
Gaussian walls spaced $80\,\rm{\mu m}$ with radial harmonic
confinement of order 15$\,$kHz geometric mean trap frequency. In
order to reach the desired final number in the range of order $10^4$
to $10^2$, the cloud is further evaporated through the h\TEM trap
for up to $50\,\rm{ms}$ (`h\TEM, trap, evap.' in Table
\ref{tab:beamtable}). Finally, the v\TEM and h\TEM beams are ramped
to the full value (`full trap' in Table \ref{tab:beamtable})
producing a mean radial trap frequency of order 40$\,$kHz. An
absorption image of a BEC produced in this fashion is shown in Fig.
\ref{fig:becfigure}.

As mentioned earlier, the h\TEM beam multiplex, the end-cap beams,
and the compensation beam have similar construction. The basic idea
is to use a multiple frequency acousto-optical modulator (AOM). This
AOM is driven with $n$ distinct radio frequencies of the form:
$V_{\rm RF} = \sum_{i=1}^n A_i\cos(\omega_i t)$, where the
amplitudes, $A_i$, and frequencies, $\omega_i$, are independently
controlled in the experiment. Each frequency produces a first order
spot \cite{Yariv}. For the h\TEM beam, these frequencies are
produced by separate digital RF synthesizers and combined with a RF
power combiner.
\begin{figure}
\includegraphics{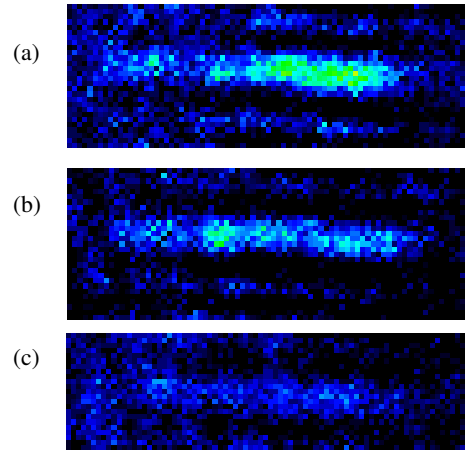}
\caption{\label{fig:compensation_figure} Absorption images of BECs
in $80\,\rm{\mu m}$ boxes. (a) $N\approx2\times 10^3$ without
compensation beam. The cloud shows a region of higher density to the
right. (b) The same as (a) but with the addition of an appropriate
compensation beam. (c) $N\approx5\times 10^2$ with compensation beam
averaged over 10 separate exposures. The color scale is different
from that in Fig \ref{fig:becfigure}. }
\end{figure}

Fig. \ref{fig:beam_figure}(b) shows a CCD image of the end-cap beam
in a general setting. Here the AOM is driven with $n=10$ independent
frequencies chosen to give an equally spaced nine site lattice. This
image demonstrates the capacity of this beam where a pair of spots
may be used to form an optical quantum dot inside the cloud. In the
case of this demonstration experiment, only a pair of spots ($n=2$)
are used, spaced by 80$\,\rm{\mu m}$ as the end caps shown in Fig.
\ref{fig:beam_figure}(c). Because the RF frequencies for this beam
are generated by separate stable voltage controlled oscillators each
with an individual RF attenuator, the number of spots, their
intensities and positions may be controlled independently on the
$10\,\rm{\mu s}$ time scale.

\begin{figure}
\includegraphics{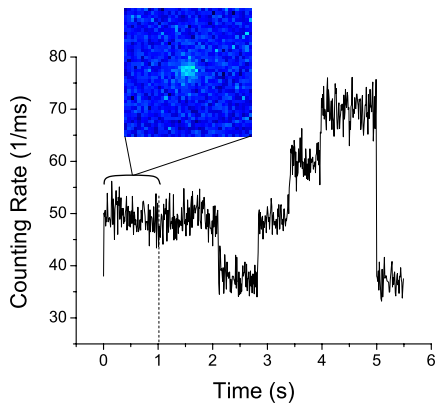}
\caption{\label{fig:SAD} Fluorescence signal for an atom transferred
from the optical trap into the small MOT at time zero. The signal
shows a single initial atom is transferred and held for 2$\,$s. The
level changes afterwards are due to random loss and background
loading. The inset is a fluorescence image of the single atom for a
1$\,$s exposure from the beginning of this run. }
\end{figure}

Although the x\TEM trap is of good quality, it does suffer from the
problem of irregular axial potential variations of order $1\,\rm{\mu
K}\cdot k_B$. This is most likely due to scattering from the
holographic plate used to produce the \TEM beams and other imperfect
optics. This irregularity is observed to break the cloud into small
sections, a phenomenon that was also observed in atom chip
experiments \cite{Leanhardt}. In either case, this is due to
potential variations on the order of the BEC chemical potential. A
compensation beam is used to fill in valleys in the axial potential.
This beam is also generated by a multiple frequency AOM, but rather
than driving it with separate RF sources, a single arbitrary
function generator is used to produce a stable frequency comb which
results in an array of spots. Fig. \ref{fig:beam_figure}(d) shows
the compensation beam driven with $n=7$ different frequencies.
Because the number of spots and their intensities are arbitrary, it
is possible in principle to create a beam with any intensity
profile. Fig. \ref{fig:beam_figure}(e) shows the compensation beam
with $n=80$ driving frequencies. The size of the structures which
may be added to the profile is limited by the spot size of the beam
(as in Table \ref{tab:beamtable}). The closeness of spacing between
the frequencies is limited by the possibility of parametric heating
of the atoms in the optical trap. This is due to beating of the
neighboring spots at their difference frequencies. Here, we operate
at a minimum difference frequency of $500\,\rm{kHz}$, which is order
$10\times$ the trap frequencies. Fig. \ref{fig:compensation_figure}
shows the effects of compensation. An uncompensated condensate shows
a larger density on the right in image (a). This variation can be
reduced with the compensation beam, as shown in Fig.
\ref{fig:compensation_figure}(b). The result is a cloud of greater
uniformity but still with some irregularities on a finer scale. This
level of compensation allows our optical trap to produce condensates
of much smaller number as shown in Fig.
\ref{fig:compensation_figure}(c) where with a better optimized
compensation profile, additional evaporation has reduced the atom
number to $5\times 10^2$. The cloud is relatively uniform which is
not possible without compensation. The small atom numbers are
cross-checked with a single atom counting system \cite{Haubrich,
Alt} which allows for independent verification of atom number. Our
atom counting setup has been integrated as to allow for atoms to be
transferred from the optical trap into a small magneto-optical trap
(MOT) so they may be accurately counted by fluorescence. Fig
\ref{fig:SAD} shows the signal of a single atom from the optical
trap. The quantized steps illustrate the capacity to determine few
atom numbers from the optical trap exactly. The level change in the
signal beyond several seconds is due to background loading and loss.
It should be emphasized that the single atom which was initially in
the dipole trap was not extracted deterministically and serves only
as a demonstration shot for the integrated detection system.

A BEC of the sort in Fig. \ref{fig:compensation_figure}(c) is a
possible initial condition for further extraction experiments such
as the quantum tweezer for atoms as proposed in ref. \cite{Diener}
where the optical quantum dot may be produced with the additional
beams in Fig. \ref{fig:beam_figure}(b). This BEC is an ideal
reservoir for single atom extraction because the mean field
splitting in the dot is of similar order to the chemical potential
of the condensate. Uniformity of the potential is currently limited
by the spot size of the compensation beam and the quality of the
absorption images used for the optimization. With improvements here,
it should be possible to obtain a cloud in the appropriate
conditions to directly study small scale Mott-insulator physics
\cite{Greiner} at the single well level and the Tonks-Girardeau (TG)
regime. The cloud may be characterized by the interaction parameter
$\gamma=m g_{\rm 1D}L/\hbar^2 N$, where $m$ is the atomic mass
\cite{Shlyapnikov, Paredes}. The case of $\gamma\ll 1$ represents
the mean-field (MF) regime whereas $\gamma\gg 1$ is a TG gas. As
calculated for our results, we have $\gamma\approx 0.08$ for Fig.
\ref{fig:becfigure} which is clearly MF, however for Fig.
\ref{fig:compensation_figure}(c) $\gamma\approx 0.5$ and for the
smallest observed condensate in our system thus far
($N\approx2.5\times 10^2$), $\gamma\approx 1$ anticipating a
borderline MF-TG regime. With a further flattened axial potential,
it should be possible to reduce the number density and push the
cloud into the TG area where it may be possible to directly study
this state. This trapping geometry has the potential for further
interrogation of these systems.

The authors would like to acknowledge support from the NSF, the R.A.
Welch Foundation, discussions with A.M. Dudarev, and comments of B.
Guti\'{e}rrez and G. Price. T.P.M. acknowledges support through the
NSF Graduate Research Fellowship and F.S. through the Alexander von
Humboldt Foundation.

\bibliography{1DBEC}

\begin{thebibliography}{19}
\expandafter\ifx\csname natexlab\endcsname\relax\def\natexlab#1{#1}\fi
\expandafter\ifx\csname bibnamefont\endcsname\relax
  \def\bibnamefont#1{#1}\fi
\expandafter\ifx\csname bibfnamefont\endcsname\relax
  \def\bibfnamefont#1{#1}\fi
\expandafter\ifx\csname citenamefont\endcsname\relax
  \def\citenamefont#1{#1}\fi
\expandafter\ifx\csname url\endcsname\relax
  \def\url#1{\texttt{#1}}\fi
\expandafter\ifx\csname urlprefix\endcsname\relax\def\urlprefix{URL }\fi
\providecommand{\bibinfo}[2]{#2}
\providecommand{\eprint}[2][]{\url{#2}}

\bibitem[{\citenamefont{Yasuda and Shimizu}(1996)}]{Yasuda}
\bibinfo{author}{\bibfnamefont{M.}~\bibnamefont{Yasuda}} \bibnamefont{and}
  \bibinfo{author}{\bibfnamefont{F.}~\bibnamefont{Shimizu}},
  \bibinfo{journal}{Phys. Rev. Lett.} \textbf{\bibinfo{volume}{77}},
  \bibinfo{pages}{3090} (\bibinfo{year}{1996}).

\bibitem[{\citenamefont{Jacksch$\;${\sl et al.}}(1999)}]{Jaksch}
\bibinfo{author}{\bibnamefont{Jacksch$\;${\sl et al.}}},
  \bibinfo{journal}{Phys. Rev. Lett.} \textbf{\bibinfo{volume}{82}},
  \bibinfo{pages}{1975} (\bibinfo{year}{1999}).

\bibitem[{\citenamefont{Diener et~al.}(2002)\citenamefont{Diener, Wo, Raizen,
  and Niu}}]{Diener}
\bibinfo{author}{\bibfnamefont{R.}~\bibnamefont{Diener}},
  \bibinfo{author}{\bibfnamefont{B.}~\bibnamefont{Wo}},
  \bibinfo{author}{\bibfnamefont{M.}~\bibnamefont{Raizen}}, \bibnamefont{and}
  \bibinfo{author}{\bibfnamefont{Q.}~\bibnamefont{Niu}},
  \bibinfo{journal}{Phys. Rev. Lett.} \textbf{\bibinfo{volume}{89}},
  \bibinfo{pages}{070401} (\bibinfo{year}{2002}).

\bibitem[{\citenamefont{Kheruntsyan$\;${\sl et al.}}(2003)}]{Kheruntsyan}
\bibinfo{author}{\bibnamefont{Kheruntsyan$\;${\sl et al.}}},
  \bibinfo{journal}{Phys. Rev. Lett.} \textbf{\bibinfo{volume}{91}},
  \bibinfo{pages}{040403} (\bibinfo{year}{2003}).

\bibitem[{\citenamefont{Paredes$\;${\sl et al.}}(2004)}]{Paredes}
\bibinfo{author}{\bibnamefont{Paredes$\;${\sl et al.}}},
  \bibinfo{journal}{Nature} \textbf{\bibinfo{volume}{429}},
  \bibinfo{pages}{277} (\bibinfo{year}{2004}).

\bibitem[{\citenamefont{Greiner$\;${\sl et al.}}(2002)}]{Greiner}
\bibinfo{author}{\bibnamefont{Greiner$\;${\sl et al.}}},
  \bibinfo{journal}{Nature} \textbf{\bibinfo{volume}{415}}, \bibinfo{pages}{39}
  (\bibinfo{year}{2002}).

\bibitem[{\citenamefont{Kinoshita$\;${\sl et al.}}(2004)}]{Kinoshita}
\bibinfo{author}{\bibnamefont{Kinoshita$\;${\sl et al.}}},
  \bibinfo{journal}{Science} \textbf{\bibinfo{volume}{305}},
  \bibinfo{pages}{1125} (\bibinfo{year}{2004}).

\bibitem[{\citenamefont{Tolra$\;${\sl et al.}}(2004)}]{Tolra}
\bibinfo{author}{\bibfnamefont{L.}~\bibnamefont{Tolra$\;${\sl et al.}}},
  \bibinfo{journal}{Phys. Rev. Lett.} \textbf{\bibinfo{volume}{92}},
  \bibinfo{pages}{190401} (\bibinfo{year}{2004}).

\bibitem[{\citenamefont{Saleh and Teich}(1991)}]{Saleh}
\bibinfo{author}{\bibnamefont{Saleh}} \bibnamefont{and}
  \bibinfo{author}{\bibnamefont{Teich}}, \emph{\bibinfo{title}{Fundamentals of
  Photonics}} (\bibinfo{publisher}{Wiley}, \bibinfo{year}{1991}).

\bibitem[{one()}]{oned_bec}
\bibinfo{note}{\small{A BEC is lower dimensional when $\mu_{\rm
  3D}\lesssim\hbar\omega_t$, where $\mu_{\rm 3D}$ is the 3D chemical potential
  and $\omega_t$ is the trapping frequency in the strongly confined
  direction(s) \cite{Shlyapnikov, Gorlitz}. This reduces the coupling parameter
  to $g_{\rm 1D}=2a_s\hbar\omega_\rho$, where $a_s\cong5.3\,$nm is the s-wave
  scattering length and $\omega_\rho$ is the geometric mean trap frequency in
  the strong directions. In the Thomas-Fermi limit, the chemical potential is
  $\mu_{\rm 1D}=g_{\rm 1D}N/L$, where $N$ is the atom number and $L$ is the
  condensate length. The maximum number of atoms in an effective 1D condensate,
  $N_{\rm max}=L/4a_s$, depends on the length of the trap rather than the
  strength of transverse confinement, which is $\cong 3500$ for us. Increased
  radial trap strength gives higher $\mu_{\rm 1D}$, for experiments of the
  Mott-insulator \cite{Greiner} or quantum tweezer \cite{Diener} sort, for
  $\,^{87}$Rb as high as $30\,$kHz is required, which is met by our optical
  trap.}}

\bibitem[{\citenamefont{Bongs$\;${\sl et al.}}(2001)}]{Bongs}
\bibinfo{author}{\bibnamefont{Bongs$\;${\sl et al.}}}, \bibinfo{journal}{Phys.
  Rev. A} \textbf{\bibinfo{volume}{63}}, \bibinfo{pages}{031602}
  (\bibinfo{year}{2001}).

\bibitem[{\citenamefont{Meyrath$\;${\sl et al.}}(2005)}]{Meyrath}
\bibinfo{author}{\bibnamefont{Meyrath$\;${\sl et al.}}},
  \bibinfo{journal}{cond-mat/0503349}  (\bibinfo{year}{2005}).

\bibitem[{\citenamefont{G\"{o}rlitz$\;${\sl et al.}}(2001)}]{Gorlitz}
\bibinfo{author}{\bibnamefont{G\"{o}rlitz$\;${\sl et al.}}},
  \bibinfo{journal}{Phys. Rev. Lett.} \textbf{\bibinfo{volume}{87}},
  \bibinfo{pages}{130402} (\bibinfo{year}{2001}).

\bibitem[{\citenamefont{Grimm et~al.}(2000)\citenamefont{Grimm,
  Weidem\"{u}ller, and Ovchinnikov}}]{Grimm}
\bibinfo{author}{\bibnamefont{Grimm}},
  \bibinfo{author}{\bibnamefont{Weidem\"{u}ller}}, \bibnamefont{and}
  \bibinfo{author}{\bibnamefont{Ovchinnikov}}, \bibinfo{journal}{Adv. At. Mol.
  Opt. Phys.} \textbf{\bibinfo{volume}{42}}, \bibinfo{pages}{95}
  (\bibinfo{year}{2000}).

\bibitem[{\citenamefont{Yariv and Yeh}(1984)}]{Yariv}
\bibinfo{author}{\bibnamefont{Yariv}} \bibnamefont{and}
  \bibinfo{author}{\bibnamefont{Yeh}}, \emph{\bibinfo{title}{Optical Waves in
  Crystals}} (\bibinfo{publisher}{Wiley}, \bibinfo{year}{1984}).

\bibitem[{\citenamefont{Leanhardt$\;${\sl et al.}}(2003)}]{Leanhardt}
\bibinfo{author}{\bibnamefont{Leanhardt$\;${\sl et al.}}},
  \bibinfo{journal}{Phys. Rev. Lett.} \textbf{\bibinfo{volume}{90}},
  \bibinfo{pages}{100404} (\bibinfo{year}{2003}).

\bibitem[{\citenamefont{Haubrich$\;${\sl et al.}}(1996)}]{Haubrich}
\bibinfo{author}{\bibnamefont{Haubrich$\;${\sl et al.}}},
  \bibinfo{journal}{Europhys. Lett.} \textbf{\bibinfo{volume}{34}},
  \bibinfo{pages}{663} (\bibinfo{year}{1996}).

\bibitem[{\citenamefont{Alt}(2002)}]{Alt}
\bibinfo{author}{\bibnamefont{Alt}}, \bibinfo{journal}{Optik}
  \textbf{\bibinfo{volume}{113}}, \bibinfo{pages}{142} (\bibinfo{year}{2002}).

\bibitem[{\citenamefont{Petrov et~al.}(2000)\citenamefont{Petrov, Shlyapnikov,
  and Walraven}}]{Shlyapnikov}
\bibinfo{author}{\bibnamefont{Petrov}},
  \bibinfo{author}{\bibnamefont{Shlyapnikov}}, \bibnamefont{and}
  \bibinfo{author}{\bibnamefont{Walraven}}, \bibinfo{journal}{Phys. Rev. Lett.}
  \textbf{\bibinfo{volume}{85}}, \bibinfo{pages}{3745} (\bibinfo{year}{2000}).

\end{thebibliography}

\end{document}